\documentclass[a4paper]{jpconf}
\usepackage{amsmath,amssymb,graphicx}

\begin{document}
\title{Dynamics of the quantum vacuum:
       Cosmology as relaxation to the equilibrium state}

\author{F.R. Klinkhamer$^1$ and G.E. Volovik$^{2,3}$}

\address{$^1$ Institute for Theoretical Physics, University of Karlsruhe,
        Karlsruhe Institute of Technology, 76128 Karlsruhe, Germany}
\address{$^2$ Low Temperature Laboratory, Aalto University,
         P.O. Box 15100, FI--00076 AALTO, Finland}
\address{$^3$ Landau Institute for Theoretical Physics RAS, Kosygina 2,
         119334 Moscow, Russia}

\ead{frans.klinkhamer@kit.edu; volovik@boojum.hut.fi
}

\begin{abstract}
The behavior of the gravitating vacuum energy density in an expanding
universe is discussed. A scenario is presented with a
step-wise relaxation of the vacuum energy density.
The vacuum energy density moves from plateau to plateau and
follows, on average, the steadily decreasing matter energy density.
The current plateau with a small positive value of the
vacuum energy density (effective cosmological constant)
may result from a still not equilibrated contribution
of the light massive neutrinos to the quantum vacuum.
\vspace*{.5\baselineskip}\newline
Preprint Nos.: arXiv:1102.3152,\,KA-TP-03-2011 
\vspace*{.5\baselineskip}\newline
Journal Ref.: J. Phys. Conf. Ser. \textbf{314} (2011) 012004
\end{abstract}

\section{Introduction}
\label{sec:Introduction}

\thispagestyle{plain}  

There are many different contributions to the vacuum energy density.
They can be separated into sub-Planckian and trans-Planckian
contributions with respect to the energy scale $E_{\rm Planck}
\equiv  [\hbar\,c^{5}/(8\pi\,G_{N})]^{1/2}
\approx 2.44\times 10^{18}\;\text{GeV}$.
The sub-Planckian contributions are described by relativistic
quantum fields propagating over a classical spacetime manifold.
The trans-Planckian contributions
come from the fundamental  microscopic degrees of freedom of
the ``deep vacuum.'' In the perfect equilibrium Minkowski
vacuum~\cite{KlinkhamerVolovik2008a}, the  trans-Planckian degrees
of freedom compensate the contribution of the sub-Planckian
quantum fields to the gravitating vacuum energy density $\rho_{\rm vac}$
(effective cosmological constant $\Lambda_{\rm eff}$).

The above discussion adopts the condensed-matter point of view,
where the effective quantum fields
emerge only at low energies and where the high-energy phenomena
are determined by the fundamental (atomic) degrees of freedom.
In a self-sustained equilibrium vacuum, the full compensation
of the vacuum energy density occurs without
fine-tuning~\cite{KlinkhamerVolovik2008a,KlinkhamerVolovik2008b}:
the microscopic degrees of freedom are automatically adjusted
to the equilibrium state.
The self-tuned nullification of the energy density
in the ground state of an arbitrary equilibrium system,
including the relativistic quantum vacuum, suggests
a possible solution~\cite{KlinkhamerVolovik2010}
of the main cosmological constant problem~\cite{Weinberg1988}.
A brief review of this so-called $q$--theory
approach to the cosmological constant problem is
presented in~\ref{sec:Resume}.

The $q$--theory approach to the cosmological constant problem
differs from those approaches which consider only relativistic
quantum fields and where the contributions of the different
quantum fields, fermionic and bosonic, compensate each other.
This fermion-boson com\-pen\-sa\-tion can be due
to supersymmetry or to a special choice of the fermionic
and bosonic content of the vacuum, relying on special relations between
the masses of the scalar, spinor,
and vector fields (see, e.g., Refs.~\cite{Zeldovich1968,
FrolovFursaev1998,Alberghi2008,Barvinsky-etal2009,Froggatt2010}).

In the perfect equilibrium state, the individual contributions
to the vacuum energy density cannot be distinguished.
But the situation changes dramatically for the dynamical vacuum
of an \underline{expanding} universe~\cite{KlinkhamerVolovik2008b}.
Cosmology, then, corresponds to the process of relaxation
towards the equilibrium vacuum state, with the vacuum energy density
dropping from its initial large value
[which can be of order $\rho_{\rm vac}\sim (E_{\rm Planck})^4\,$]
to its present small value [$\rho_{\rm vac}\approx(2\;\text{meV})^4\,$].
During the expansion of the Universe,
the energy hierarchy of the different contributions
to the vacuum energy density from the different quantum fields
gives rise to a time sequence, since each epoch is characterized
by one dominant contribution to the vacuum energy density and, thus,
by one particular value of the effective cosmological constant.
This suggests a \underline{step-wise relaxation}
of the effective cosmological constant,
which, on average, follows the matter energy density.

The contribution of a quantum matter field  to the vacuum energy
density is given by the following expression (see, e.g.,
Refs.~\cite{Zeldovich1968,FrolovFursaev1998,Alberghi2008,Barvinsky-etal2009,
Froggatt2010,Visser2002,Akhmedov,OssolaSirlin}):
\begin{equation}
\rho_{\rm vac}^\text{\,(matter)}(M,\,E_{\rm uv})
=c_4\, (E_{\rm uv})^4 + c_2\,(E_{\rm uv})^2\,M^2 +
c_0\,M^4\,\ln\big[\,(E_{\rm uv})^2\big/M^2\,\big]
+\text{O}\big(M^4\big)~,
\label{General}
\end{equation}
where $M$ is the mass of the matter field considered and $E_{\rm uv}$
is the ultraviolet energy cutoff, which is typically assumed
to be of the order of the energy scale $E_{\rm Planck}$ defined above.
Different regularization schemes have been
suggested, in order to obtain the coefficients $c_{n}$
appearing in \eqref{General}. Here, and in the following,
natural units are used with $\hbar=c=k_{B}=1$.

Cosmological phase transitions or crossovers
lead to time-varying masses of the
fermionic and bosonic fields and, thereby, affect the contributions
of these fields to the effective cosmological constant.
In this article, the focus will be on the fermionic degrees of freedom,
whose number is larger than the one of the bosonic degrees of freedom
in the standard model of elementary particle physics
(cf. Sec.~19.3.2 of Ref.~\cite{PDG2010}).

It is natural to suppose that, for higher temperatures,
the number of fermionic quantum fields with nonzero masses is smaller,
thereby removing the corresponding negative contributions
to the vacuum energy density
(see Secs.~\ref{sec:Fermion propagator}--\ref{sec:Fermion-contrib-QCD}
and \ref{sec:Neutrino-mass-cutoff} for details, in particular,
for the reason of having a \emph{negative} contribution).
As a result, the vacuum energy density from fermionic mass contributions
is typically large for high temperature and small for low temperature.
In an expanding universe, this vacuum energy density
(effective cosmological constant) then
relaxes in a step-wise manner, following the appearance
of masses of the fermionic matter fields
(see Secs.~\ref{sec:Fermion-contrib-neutrinos}--\ref{sec:Conclusion}
for further discussion). The step-wise relaxation
continues until the vacuum energy density reaches
the zero value of the perfect equilibrium state
or until the vacuum energy density is frozen at a stage where the
relaxation process becomes ineffective.

Several scenarios have been proposed for this final (frozen) stage and the
corresponding remnant vacuum energy density. These scenarios rely, for
example, on new TeV--scale electroweak
physics~\cite{ArkaniHamed-etal2000,KlinkhamerVolovik2009b,Klinkhamer2011} or
nonperturbative QCD
dynamics~\cite{Schutzhold2002,KlinkhamerVolovik2009a,Klinkhamer2010,
UrbanZhitnitsky2009,Holdom2011} with a remnant vacuum energy density given
by, respectively, $\rho_{{\rm vac},\infty}\sim (E_{\rm ew})^8/(E_{\rm
Planck})^4$ or $\rho_{{\rm vac},\infty}\sim (E_{\rm QCD})^6/(E_{\rm
Planck})^2$. In Sec.~\ref{sec:Fermion-contrib-neutrinos}, we add a scenario
related to neutrino mass $M_{\nu}$, which may give a remnant vacuum energy
density of order $(M_{\nu})^4$; see, e.g.,
Refs.~\cite{Capolupo-etal2007,BauerSchrempp2008,Chitov-etal2011} for related
discussions. All these scenarios do not exclude each other. Different
contributions to the vacuum energy density produce successive plateaux in the
process of the relaxation of the effective cosmological constant and the one
which survives (or is frozen) is responsible for the current plateau of the
effective cosmological constant at the meV--scale level.

\section{Fermion propagator in an interacting system
         and vacuum energy density $\boldsymbol{\rho_\text{vac}}$}
\label{sec:Fermion propagator}

As a start, consider the contribution of a fermionic quantum field
to the vacuum energy density. The discussion will be based on
certain properties of the mass function in the fermion propagator.
This fermion propagator must apply to
the virtual fermions of the quantum vacuum
(the ground state in condensed-matter-physics language)
which is a dense interacting system with Planck-scale energies.

The general form of the Green's function of a fermionic particle is
\begin{equation}
G(p)  =\frac{Z(p^2)}{i\,\gamma^{\mu}\, p_{\mu}+ M(p^2)} \;,
\label{Dirac-propagator}
\end{equation}
for a Euclidean metric,
$p^2\equiv p_{\mu} \, p_{\nu}\, \delta^{\mu\nu} = |{\bf p}|^2 +(p_0)^2$,
and appropriate Dirac matrices,
$\gamma^{\mu} \gamma^{\nu} +\gamma^{\nu} \gamma^{\mu} =2\,\delta^{\mu\nu}$.
The mass function $M(p^2)$ in \eqref{Dirac-propagator} must disappear
at large $p^2$, where interaction effects can be neglected.
The simplest form  of $M(p^2)$ satisfying that condition is
 \begin{equation}
 M(p^2)= \frac{a}{b + p^2} \;,
\label{eq:M-analytic-Ansatz}
\end{equation}
in term of parameters $a>0$ and $b\geq 0$ with mass dimensions
3 and 2, respectively.

This two-parameter \textit{Ansatz}
reflects two important physical quantities of the mass function,
which depend on the interaction strength characterizing the system.
First, the parameter $b^{1/2}$ from model \eqref{eq:M-analytic-Ansatz}
gives the energy scale of the effective ultraviolet cutoff
for mass-dependent effects.
Second, the ratio of the parameters $a$ and $b$ represents the mass
parameter at zero momentum,
$M(0)=a/b=\big(a\,b^{-3/2}\big)\,b^{1/2}$.
The fermion mass $\overline{M}$ at the pole of the Green's function
(i.e., on mass shell, temporarily reverting to a Lorentzian metric)
is determined by the equation
$\overline{M}^{\,2}\,\big(b-\overline{M}^{\,2}\,\big)^2=M^2(0)\,b^2$.
For $a^2\ll b^3$, the on-shell
mass is simply given by $\overline{M}\approx M(0)=a/b$.

For charged leptons, the natural value of the cutoff
is the electroweak energy scale, $b\sim (E_{\rm ew})^2$.
For neutrinos, the cutoff may be comparable with the
neutrino mass itself
(a heuristic argument based on the see-saw mechanism
is given in \ref{sec:Neutrino-mass-cutoff}).
For the analytic \textit{Ansatz} \eqref{Dirac-propagator},
this gives $a^2\sim b^3$ and the on-shell
neutrino mass $\overline{M}_{\nu}$ typically differs from $M_{\nu}(0)$
but is still of order $M_{\nu}(0)$.

The fermionic contribution to the vacuum energy density
is obtained from the following
Euclidean effective action:
\begin{equation}
S_{E}= -  V_{4}\;\int  \frac{d^4 p}{(2\pi)^4} \;
          \text{tr} \,   \ln \big[G(p)\,\widetilde{m}\,\big]\,,
\label{Action}
\end{equation}
where $V_{4}$ is the Euclidean spacetime volume considered
and $\widetilde{m}$ a fixed reference mass. The zeroth-order term
in the gradient expansion gives the contribution
of the fermionic field to the cosmological constant term,
\begin{equation}
S=-\int d^{4}x\,\sqrt{-\text{det}[g(x)]}\;\;
  \rho_{\rm vac}^\text{\,(fermion)} + \ldots\,,
\label{Action-leading-term}
\end{equation}
where, now, the action has been given
for an arbitrary Lorentzian metric $g_{\mu\nu}(x)$
and the conventions of Ref.~\cite{KlinkhamerVolovik2010}
have been used.

The fermionic contribution to the cosmological constant can be
calculated by introducing the vierbein field in \eqref{Action},
$\gamma^a\, e^{\mu}_a\, p_{\mu}$, and
taking the functional derivative of $S_{E}$ with respect to $e^{\mu}_a$.
Consider the contribution from the denominator of the
Green's function \eqref{Dirac-propagator}.
In order to obtain a simple estimate of the mass effects,
take the derivative of \eqref{Action} with respect to $a$,
 \begin{equation}
\frac{d\, S_{E}}{d\, a}
\sim
\int  \frac{d^4 p}{(2\pi)^4}\;\text{tr}\big[\, (b+ p^2)^{-1}\,G\, \big]
\sim
\int  \frac{d^4 p}{(2\pi)^4}\; \frac{a}{a^2 + p^2\,(b+ p^2)^2}\;,
   \label{ActionDerivative}
\end{equation}
leaving out $V_{4}$ for the sake of brevity.
The case $b^3\lesssim a^2$ gives $dS_{E}/da \sim a^{1/3}$ and $S_{E} \sim a^{4/3}$.
The case $a^2 \lesssim b^3$ gives $S_{E} \sim a^2/b$.
Combined, there is thus the following simple estimate of
the fermion contribution to the vacuum energy density:
\begin{equation}
\rho_{\rm vac}^\text{\,(fermion)}
\sim
\left\{\begin{array}{l}
  -a^{4/3}\,
  \;\;\;\; \text{for}\;\;\;\;
  b^3 \lesssim a^2
  \,, \\[1mm]
  -a^2/b
  \;\;\;\; \text{for}\;\;\;\;
  b^3 \gtrsim a^2 \,,
\end{array}\right.
\label{rho_V-fermion}
\end{equation}
where the overall minus sign will be verified independently
in the next section.

\section{Fermion contributions to $\boldsymbol{\rho_\text{vac}}$ at
         the electroweak energy scale}
\label{sec:Fermion-contrib-EW}

In the electroweak standard model, the natural choice for the
Green's function of quarks and
leptons with masses $M\lesssim E_{\rm ew}$
gives parameters $b\sim (E_{\rm ew})^2$ and
$a=M(0)\,b \lesssim (E_{\rm ew})^3$ for the
model mass function \eqref{eq:M-analytic-Ansatz}.
According to \eqref{rho_V-fermion},
the contribution to the vacuum energy density
from light fermions with $M\ll E_{\rm ew}$ is
of order $M^2\, (E_{\rm ew})^2$
and the one of heavy fermions  with $M\sim E_{\rm ew}$
of order $(E_{\rm ew})^4$.
Hence, the general expectation is that the contribution of a
charged fermionic field with mass $M\lesssim E_{\rm ew}$
to the vacuum energy density is
 \begin{equation}
\rho_{\rm vac}^\text{\,(fermion)}(M) \sim - M^2 \,(E_{\rm ew})^2\,,
\label{MassContribution}
\end{equation}
taking over the minus sign from \eqref{rho_V-fermion}.

The previous estimates \eqref{rho_V-fermion} and
\eqref{MassContribution} are obtained from an analysis of
the denominator of the Green's function \eqref{Dirac-propagator},
but the numerator gives similar results.
The residue $Z(p^2)$ approaches unity as $p^2\to \infty$.
For finite values of $p^2$, however, the residue $Z(p^2)$ is
again determined by two energy scales, $M(0)$ and $E_{\rm ew}$.

Estimate \eqref{MassContribution}, including the minus sign,
can also be obtained by direct calculation of the
mass contribution to the zero-point energy from a
spin--$\textstyle{\frac{1}{2}}$ Dirac field:
\begin{equation}
\int^{(E_{\rm cutoff})}\, \frac{d^3 \mathbf{p}}{(2\pi)^3}
\; \Big( -\sqrt{|\mathbf{p}|^2+ M^2}+ |\mathbf{p}|\, \Big)
\sim - M^2\, (E_{\rm cutoff})^2\,,
\label{QuarkContribution2}
\end{equation}
where $E_{\rm cutoff}$ is the ultraviolet cutoff of this
quadratically divergent integral (see, e.g., Ref.~\cite{Zeldovich1968}).
Since the Green's function of a massive standard-model fermion differs
from the Green's function of a massless fermion only at energies
below $E_{\rm ew}$, it is the electroweak scale $E_{\rm ew}$ which
provides the natural ultraviolet cutoff in \eqref{QuarkContribution2}
rather than the Planck scale: $E_{\rm cutoff}=E_{\rm ew}$.

A Planck-scale cutoff $E_{\rm cutoff}=E_{\rm Planck}$
in \eqref{QuarkContribution2}
would be relevant only if the mass of the fermion were fundamental
or generated at the Planck energy scale,
i.e., with $b\sim (E_{\rm Planck})^2$ in the mass function
\eqref{eq:M-analytic-Ansatz}. However, if
the Green's function of a massive fermion differs from
the Green's function of a massless fermions only
at energies of order of mass scale (as discussed
in the fourth paragraph of Sec.~\ref{sec:Fermion propagator}),
then the cutoff in \eqref{QuarkContribution2}
is determined by the fermion mass itself,
leading to $ \rho_{\rm vac}^\text{\,(fermion)}(M)\sim - M^4$.
We suggest that the last estimate holds for the neutrino
contribution (see Sec.~\ref{sec:Fermion-contrib-neutrinos}
for possible cosmological implications).

\section{Fermion contributions to $\boldsymbol{\rho_\text{vac}}$
         at the QCD energy scale}
\label{sec:Fermion-contrib-QCD}

For the quarks of quantum chromodynamics (QCD),
the  natural choice  of parameters in model \eqref{eq:M-analytic-Ansatz}
is $a\sim (E_{\rm QCD})^3$ and  $b=0$.
Such a choice effectively corresponds to the
phenomenon of confinement, which leads to
a singular behavior of the effective quark mass
in the infrared limit $p \to 0$.
Actually, this Green's function at $p_0=0$  has been used
in Refs.~\cite{KlinkhamerVolovik2009a,Klinkhamer2010}
to justify  the presence of a running  term
$|H(t)|\,(E_{\rm QCD})^3$ in the vacuum energy density
[here, $H(t)$ is the Hubble parameter of the expanding universe]
and to calculate the corresponding remnant vacuum energy density
$\rho_{{\rm vac},\infty}\sim (E_{\rm QCD})^6/(E_{\rm Planck})^2$.

However, the justification of the term $|H|\,(E_{\rm QCD})^3$ is still
problematic. Model \eqref{eq:M-analytic-Ansatz} with  $a\sim (E_{\rm QCD})^3$
and  $b=0$ gives, in first approximation \eqref{rho_V-fermion}, a large
vacuum energy density, $|\rho_{\rm vac}^\text{\,(fermion)}|\sim (E_{\rm
QCD})^4$. With $a>0$, there is no infrared divergence in the integral
\eqref{ActionDerivative} for $b=0$ and the suggested $|H|\,(E_{\rm QCD})^3$
correction does not appear. Still, this does not completely rule out a term
of order $(E_{\rm QCD})^6/(E_{\rm Planck})^2$ in the vacuum energy density: a
more elaborate model of the QCD vacuum energy density is needed, which
properly takes into account the phenomenon of confinement. There is a hint
from lattice simulations, that such a $(E_{\rm QCD})^6/(E_{\rm Planck})^2$
term is indeed possible~\cite{Holdom2011}.

\section{Fermion contributions to $\boldsymbol{\rho_\text{vac}}$
         at the mass scale of light neutrinos}
\label{sec:Fermion-contrib-neutrinos}

At finite temperature $T$, the mass function $M(p^2)$
in \eqref{Dirac-propagator} can be expected to depend also on $T$.
Moreover,  temperature may provide the relevant energy scale
for the ultraviolet cutoff in \eqref{QuarkContribution2},
$E_{\rm cutoff}\sim T$. As a result, the mass-dependent contribution
becomes part of the thermal energy rather than the vacuum energy.
This suggests a natural form of energy exchange between
thermal fermions and vacuum fermions,
i.e., between matter and vacuum. In any case,
it can be expected that, with increasing temperature,
the mass-dependent contributions to the vacuum energy density
will be reduced or even disappear.

\begin{figure}[t] 
\centering 
\includegraphics[width=0.95\textwidth]{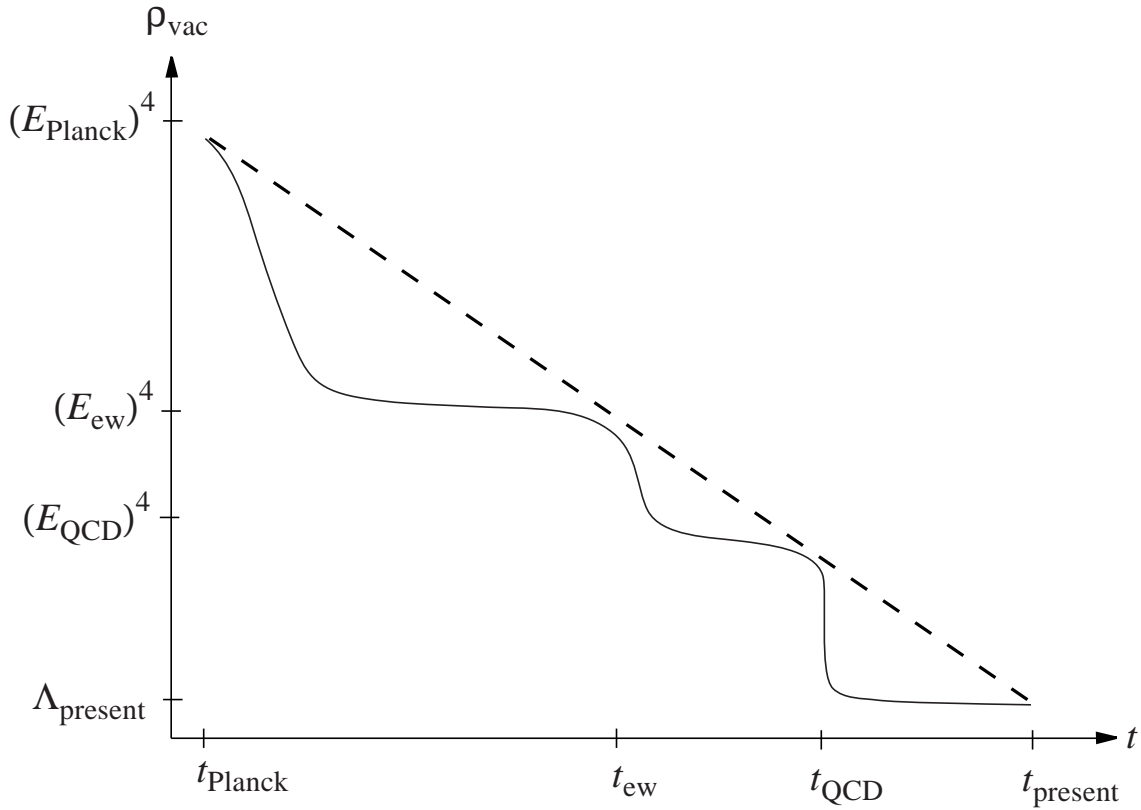} 
\vspace*{-4mm} 
\caption{Sketch (as an approximate double-log plot)
of the relaxation of the vacuum energy density
during the evolution of the Universe.
{\it Dashed curve}:  relaxation according to the relation
$<\rho_{\rm vac}(t)>  \;\sim\, (E_{\rm Planck})^2/t^2$
from Ref.~\cite{KlinkhamerVolovik2008b}.
Only the envelope of the ultrafast (Planck-scale) oscillations
of $\rho_{\rm vac}(t)$ is shown, that is, $<\rho_{\rm vac}(t)>$.
The huge vacuum energy density $\rho_{\rm vac}\sim (E_{\rm Planck})^4$
at the initial  moment $t_{\rm in}\sim t_{\rm Planck}$
relaxes to a zero value as $t\rightarrow \infty$.
According to this estimate,
the vacuum energy density at the present time passes through
a value close to the one of the observed cosmological ``constant'':
$\rho_{\rm vac}(t_{\rm present})$
$\sim$
$(E_{\rm Planck})^2/t_{\rm present}^2$
$\sim$
$\Lambda_{\rm present}^\mathrm{(obs)}$
$\approx$ $(2\;\text{meV})^4$.
{\it Full curve}:
dissipative processes and cosmological phase transitions
lead to a modified behavior, with a
step-wise decrease of the vacuum energy density towards zero.
The vacuum energy density follows the matter energy density on average,
but locally, on each plateau, the vacuum behaves as a medium with
equation-of-state parameter $w\approx -1$.
}
\label{fig:Relaxation}
\vspace*{0cm}
\end{figure}

As the temperature of the adiabatically expanding universe
decreases, more and more fermions become massive and add
extra negative contributions \eqref{QuarkContribution2}
to the vacuum energy density.
This implies that the total vacuum energy density
$\rho_{\rm vac}$ decreases with time and that this change occurs
in a step-like manner (Fig.~\ref{fig:Relaxation}),
following, on the whole, the matter energy density.

According to this scenario, one of the plateaux
in  $\rho_{\rm vac}(t)$  for $t\sim \overline{t}$
results from the lack of the
neutrino-mass contribution to the vacuum energy density,
$ \rho_{\rm vac}^\text{(neutrino)}(M_{\nu})\sim - (M_{\nu})^4$.
The lack of a negative neutrino contribution
to the vacuum energy density corresponds, of course,
to a positive value of $\rho_{\rm vac}(\overline{t})$
at the time scale $\overline{t}$ considered. Contributions of
fermionic matter fields with larger vacuum energy densities
have already been released at higher temperatures and have been relaxed
due to the fast oscillations of the microscopic $q$--type
fields (see \ref{subsec:Resume-Dynamics}).

Restrict the discussion temporarily to one neutrino flavor. Then,
the neutrino contribution to the vacuum energy density can only be
released after the temperature $T_{\nu}$ of the relic neutrino drops
below the neutrino mass $M_{\nu}$
[which happens at a redshift of order
$M_{\nu}/(3\times 10^{-4}\,\text{eV})\sim 10^2$
for $M_{\nu} \sim 0.05\;\text{eV}\,$].
It is, however, very well possible that the contribution
from the light neutrino is
\underline{still not released} at the present moment, because
of the lack of an efficient equilibration mechanism between
neutrino matter and neutrino vacuum in the current epoch.
In that case, the excess of the vacuum energy density
remains frozen at a \underline{positive} value of order
$(M_{\nu})^4$. The lowest plateau in Fig.~\ref{fig:Relaxation}
can be expected to extend far beyond the present age of the Universe.

Turning to three neutrino flavors
with substantial (near-maximal) mixing~\cite{PDG2010},
a heuristic argument suggests the following expression for
the frozen value of the vacuum energy density:
\vspace*{-2mm} 
\begin{equation}
\rho_{\rm vac}(t_\text{present})
\;\stackrel{?}{=}\;
c_\nu\,\Big(\text{tr}\, \big[(\widehat{M}_{\nu})^2\,\big]\Big)
\;\Big(\text{det}\,\big[(\widehat{M}_{\nu})^{2}\,\big]\Big)^{1/3}\;,
\vspace*{-2mm} 
\label{NeutrinoContribution}
\end{equation}
where $\widehat{M}_{\nu}$ is the $3\times 3$ neutrino mass matrix
with eigenvalues $\{m_{\nu 1},\, m_{\nu 2},\, m_{\nu 3} \}$ and
the positive coefficient $c_\nu$  is assumed to be of order 1.
Additional terms, possibly depending on the mixing angles,
can be expected in \eqref{NeutrinoContribution}.
Note that neutrino contributions
to the effective cosmological constant have been discussed
in a number of recent articles (see, e.g.,
Refs.~\cite{Capolupo-etal2007,BauerSchrempp2008,Chitov-etal2011}).

Given the measured value $\rho_{\rm
vac}(t_\text{present})\approx(2.3\;\text{meV})^4$ for the left-hand side of
\eqref{NeutrinoContribution} and the neutrino-oscillation
data~\cite{PDG2010}, there are three equations for the three neutrino masses.
Take the neutrino-oscillation input to be $(m_{\nu 3})^2-(m_{\nu 1})^2 =
\pm\, 2.4\times 10^{-3}\;\text{eV}^2$ and $(m_{\nu 2})^2 - (m_{\nu 1})^2 =
7.7 \times 10^{-5}\;\text{eV}^2$. Then, formula \eqref{NeutrinoContribution}
with $c_\nu=1$ gives the following neutrino mass spectra:
\vspace*{-2mm} 
\begin{equation}
\big(m_{\nu 1},\, m_{\nu 2},\, m_{\nu 3} \big)\;\Big|^{\,(c_\nu=1)}
\;\stackrel{?}{=}\;
\left\{\begin{array}{l}
\big(2.793 \times 10^{-6}\,,\; 8.775\,,\;  48.99\big)\times \text{meV}\,,\\[1mm]
\big(48.99\,,\;  49.77\,,\; 1.783\times 10^{-7}\big)\times \text{meV}\,.
\end{array}\right.
\vspace*{-2mm} 
\label{eq:neutrino-mass-spectra}
\end{equation}
These neutrino masses (close to the minimum values needed
for the neutrino oscillations) lie outside the reach
of the KATRIN tritium beta-decay detector with a
0.2--eV design sensitivity~\cite{KATRIN-LOI2001}.
The same conclusion holds for the neutrino mass spectra
resulting from formula \eqref{NeutrinoContribution}
with $c_\nu \geq 10^{-8}$ [the $c_\nu = 10^{-8}$
spectra for both ``normal'' and ``inverted'' mass hierarchies
have $m_{\nu 1} \sim m_{\nu 2}\sim m_{\nu 3} \sim 0.2\;\text{eV}$].
See Sec.~\ref{sec:Discussion} for additional comments.

\section{Discussion}
\label{sec:Discussion}

The main message of this article is that there is a
hierarchy of different contributions
to the vacuum energy density from different matter fields.
In the context of an expanding universe, these contributions
are step-by-step released with decreasing temperature of the
Universe. Note that the few big steps of the
vacuum energy density in Fig.~\ref{fig:Relaxation} are only schematic,
there may be many additional small steps.
Some of these (small) steps of the vacuum energy density
may even increase the vacuum energy density,
if they result from the mass effects of bosonic fields
[corresponding to \eqref{QuarkContribution2} multiplied by $-1$].
But, as mentioned in Sec.~\ref{sec:Introduction},
the present article focusses on the mass effects of fermionic
fields, which typically give decreasing steps of the
vacuum energy density as the Universe expands and cools.

The vacuum energy density roughly follows
the matter energy density, but in a stepwise manner.
This suggests a possible solution of the cosmic coincidence
puzzle (cf. Ref.~\cite{ArkaniHamed-etal2000}), namely,
why the observed cosmological ``constant'' is precisely
of the order of the \emph{present} matter energy density.
Indeed, the step-wise relaxation
of the vacuum energy density may solve the cosmic coincidence puzzle,
because, during each epoch, the effective cosmological constant
is related to the energy scale
characterizing the given epoch and, thus, to the matter energy density.

It has been shown~\cite{KlinkhamerVolovik2008b}
that the huge initial vacuum energy density may continuously
relax to a zero value as the age of the Universe goes to infinity.
Moreover, it was found that this relaxing vacuum energy density
passes through the observed numerical value
($\rho_{\mathrm{vac},0} \sim 10^{-11}\;\text{eV}^{4}$)
for an age of the model universe of the order of the observed
value ($t_0 \sim 10\;\text{Gyr}$);
see, in particular, Eq.~(5.16) of Ref.~\cite{KlinkhamerVolovik2008b}.
This scenario gives a reasonable estimate
of the present value of the vacuum energy density
but contradicts the more detailed astronomical observations which
are in favor of a genuine cosmological
\emph{constant}, at least, in the current epoch.
The reason for the failure of the continuous-relaxation scenario
may be that it is oversimplified, since it does not take into
consideration the processes of radiation, dissipation, and
cosmological phase transitions or crossovers.
The new scenario with a step-wise decrease of the
vacuum energy density may remove this discrepancy:
the vacuum energy density follows the matter energy density on average,
but locally, on each plateau, behaves as a medium with equation-of-state
parameter $w\approx -1$.

The overall picture for the evolution of the vacuum energy
density is, then, as follows.
The initial relaxation dynamics of the vacuum energy
density is dominated by microscopic (trans-Planckian) degrees of
freedom~\cite{KlinkhamerVolovik2008b},
leading to $1/t^2$ decay from ultrafast oscillations
or to exponential decay if radiation of
matter fields or gravitational waves is taken into account.
(This behavior is similar to that of Starobinsky
inflation~\cite{Starobinsky1980,Vilenkin1985},
where the role of the microscopic degrees of freedom is played
by heavy Planck-mass fields.)
For later times, probably starting at the electroweak epoch,
the vacuum dynamics and the energy exchange between vacuum and
matter is dominated by the low-energy contributions of quantum
fields and can be described by standard-model physics.
The dynamics of the trans-Planckian degrees of freedom
(e.g., the dynamics of the vacuum variable $q$ from
Ref.~\cite{KlinkhamerVolovik2008b}) becomes irrelevant
at late times, providing only a tiny response
to perturbations occurring during the
crossovers~\cite{KlinkhamerVolovik2009b,Klinkhamer2011}.
The microscopic degrees of freedom have already played their main
role: to fully compensate the vacuum energy density of quantum fields
in the final equilibrium state, that is, the Minkowski vacuum.

In order to clarify this point about the main role of the microscopic
degrees of freedom, consider, for example, the large vacuum energy
density released by an intermediate cosmological phase transition.
It is assumed that the chemical potential $\mu$
for the microscopic degrees of freedom takes its natural
value $\mu_0$, which corresponds to the true
quantum vacuum of a global equilibrium
(in the cosmological context, the final Minkowski state
with massive fermions at zero temperature),
and not a different value $\widetilde{\mu} \ne \mu_0$,
which would correspond to a nonequilibrium state
(in the cosmological context, the intermediate false-vacuum
state with massless fermions at temperatures above the phase
transition temperature considered).
This value $\mu_0$ either is fixed globally as an integration constant
being conserved throughout the history of the
Universe~\cite{KlinkhamerVolovik2008a,KlinkhamerVolovik2008b}
or emerges dynamically from an attractor-type
solution~\cite{KlinkhamerVolovik2010}.
This is the reason why the full curve in Fig.~\ref{fig:Relaxation}
(taking dissipative effects to have been operative
at $t \gtrsim t_\text{Planck}$)
does not approach a zero value of the vacuum energy density
at times $t_\text{Planck} \ll t \ll t_\text{ew}$
but a finite positive value.
Later, this vacuum energy density is reduced (released)
at the transition to the next plateau.

If the vacuum energy density in the current epoch results from
the lack of the neutrino contribution due to non-equilibration
(as discussed in Sec.~\ref{sec:Fermion-contrib-neutrinos})
and formula \eqref{NeutrinoContribution} holds with $c_\nu \sim 1$,
then the 0.2--eV sensitivity
of the KATRIN tritium beta-decay detector~\cite{KATRIN-LOI2001}
does not suffice to detect the predicted nonzero (electron-type)
neutrino mass $\lesssim 0.05\;\text{eV}$
from \eqref{eq:neutrino-mass-spectra}.

If, on the other hand, KATRIN does find a neutrino mass
$\gtrsim 0.4\;\text{eV}$, then this implies
that the neutrino contribution to the vacuum energy density
\eqref{NeutrinoContribution}
[\,assuming $c_\nu \gtrsim 10^{-6}$\,]
exceeds by several orders of magnitude the observed value of
the current effective cosmological constant
($\rho_{\mathrm{vac},0} \approx 3 \times 10^{-11}\;\text{eV}^{4}$)
and that this part of the vacuum energy density
cannot be responsible for the presently
observed value $\rho_{\mathrm{vac},0}$.
In this case, the neutrino contribution to $\rho_{\rm vac}(t)$
must have dominated in one of the previous epochs
and must have been released by now.

Assuming that the neutrino contribution has already equilibrated
(as would be suggested by a positive result from KATRIN
with $M_\nu \gtrsim  0.4\;\text{eV}$),
the remnant vacuum energy density $\rho_{{\rm vac},\infty}$
must be associated with other possible contributions
such as $(E_{\rm ew})^8/(E_{\rm Planck})^4$
from Refs.~\cite{KlinkhamerVolovik2009b,Klinkhamer2011}
or $(E_{\rm QCD})^6/(E_{\rm Planck})^2$
from Refs.~\cite{KlinkhamerVolovik2009a,Klinkhamer2010}.
However, the last two explanations of $\rho_{{\rm vac},\infty}$
may fare differently when compared with astronomical
observations: the QCD explanation corresponds to a modified gravity
theory and may already be ruled out by the astronomical observations,
whereas the electroweak explanation corresponds to a
standard $\Lambda$CDM model (now, with a calculated value
of $\Lambda$) and appears to fit the current astronomical data well.
Still, the electroweak explanation requires
nonstandard physics at the TeV energy scale, which remains
to be confirmed by particle-collider experiments.

All this also suggests that, in the past, there were additional
plateaux in the effective cosmological constant
associated with, for example, the mass $M_e$ of the electron,
the masses $M_{u,\,d}$ of the light (stable) quarks,
the QCD energy scale, and the electroweak energy scale.
These contributions to $\rho_{\rm vac}$ may be of order
$(M_e)^2 \,(E_{\rm ew})^2$,
$(M_{u,\,d})^{2}\,(E_{\rm ew})^2$,
$(E_{\rm QCD})^4$, and
$(E_{\rm ew})^4$.

\section{Conclusion}
\label{sec:Conclusion}

It has been argued, in this article, that
cosmology can be viewed as the relaxation of the Universe towards
the equilibrium vacuum state, with the vacuum energy density dropping
from its initial Planck-scale value to its present meV--scale value.
The initial relaxation dynamics of the vacuum is dominated by microscopic
(trans-Planckian) degrees of
freedom~\cite{KlinkhamerVolovik2008a,KlinkhamerVolovik2008b},
leading to $1/t^2$ decay of the average
vacuum energy density (or to exponential decay from
dissipative effects).

The dynamics of the vacuum energy density at later times is
governed by contributions to the vacuum energy density from the
relativistic quantum fields of the standard model.
The energy hierarchy of the contributions of different quantum
fields to the vacuum energy density
leads to a step-wise relaxation with time of the
effective cosmological constant (Fig.~\ref{fig:Relaxation}),
which, on average, follows the matter energy density. Such a
step-wise behavior of the vacuum energy density may solve the cosmic
coincidence puzzle~\cite{ArkaniHamed-etal2000},
because, in each epoch, the effective cosmological constant is related
to the energy scale characterizing the given epoch and, thus,
to the matter energy density. There are
several scenarios for the origin of the latest plateau
of the effective cosmological constant,
including a possible contribution \eqref{NeutrinoContribution}
from the masses of the neutrinos
(more precisely, the positive effective cosmological constant
would result from the lack of the negative mass-effect contribution
of the neutrino fields, if these fields of the quantum vacuum still
have not reached their equilibrium state).

Returning to a crucial point,
the hierarchical structure of the quantum vacuum implies a complicated
spectral function of the vacuum energy
density. (The spectral
function of the vacuum energy density has been introduced by Zeldovich
\cite{Zeldovich1968} and its relation to the standard model of
elementary particle physics has been discussed in, e.g.,
Refs.~\cite{Alberghi2008,Kamenshchik2007,Volovik2009}.)
The individual contributions to the spectral function
of the vacuum energy density
cannot be resolved in the static Minkowski vacuum,
since, in equilibrium, the contributions of the bosonic and fermionic
quantum fields to the low-energy part of  the spectral function are
compensated by the trans-Planckian part of the
spectrum~\cite{Volovik2009}.
The trans-Planckian degrees of freedom, which are responsible
for the automatic nullification of the cosmological constant
in the equilibrium Minkowski vacuum, play also an important role
in the dynamics at the early (Planckian) stage of
expansion of the Universe.
At later stages, the dynamics of the quantum vacuum
is primarily determined by the low-energy tail of the spectral
function. The various hierarchies of low-energy scales give rise
to different plateaux in the vacuum energy density, each of
which resembles a genuine cosmological constant
but has a dynamic origin nevertheless.

\ack

FRK thanks the Perimeter Institute for Theoretical Physics, Canada,
and the Department of Physics, University of Toronto, Canada,
for their hospitality in September, 2010.
The work of GEV is supported in part by the Academy of Finland,
Centers of Excellence Program 2006--2011.

\appendix
\section{R\'{e}sum\'{e} of $\mathbf{q}$--theory}
\label{sec:Resume}

One route to understanding the gravitational effects
of the vacuum energy density goes under the name of ``$q$--theory''
~\cite{KlinkhamerVolovik2008a,KlinkhamerVolovik2008b,KlinkhamerVolovik2010}.
The idea is to consider the \underline{macroscopic} equations of a
conserved \underline{microscopic} variable $q$,
whose precise nature need not be known.
The goal of this appendix is to explain the basic logic
of $q$--theory as an
approach to the cosmological constant problem~\cite{Weinberg1988}.

\vspace*{-2mm} 
\subsection{Statics}
\label{subsec:Resume-Statics}

For a special type of vacuum variable $q$
introduced in Ref.~\cite{KlinkhamerVolovik2008a},
the vacuum energy density $\rho_\text{vac}(q)$ entering the
gravitational field equations has been found
to differ from the vacuum energy density $\epsilon(q)$
appearing in the action:
\vspace*{-2mm} 
\begin{equation}
\rho_\text{vac}(q)=
\epsilon(q) -q\;\frac{d \epsilon(q)}{d q}
\equiv
\epsilon(q) -q\;\epsilon^\prime(q)\;.
\vspace*{-2mm} 
\label{rhoV-GibbsDuhem}
\end{equation}
Here, $q$ is a microscopic variable describing
the physics of the deep (ultraviolet) vacuum, but its thermodynamics
and dynamics are governed
by macroscopic equations, because $q$ is a \underline{conserved} quantity.
This quantity $q$ is similar to the mass density in liquids, which
describes a microscopic quantity -- the number density of atoms -- but obeys
the macroscopic equations of hydrodynamics, because of particle-number
conservation.

However, different from known liquids,
the quantum vacuum is \underline{Lorentz invariant}.
The quantity $q$ must, therefore, be Lorentz invariant, at least, in the
equilibrium state,
\vspace*{-2mm} 
\begin{subequations}\label{q0-rhovacq0}
\begin{equation}
q=q_0=\text{constant}\,.
\vspace*{-2mm} 
\label{q0}
\end{equation}
The variable $q$ naturally allows for a vanishing gravitating vacuum
energy density (cosmological constant)
\vspace*{-2mm} 
\begin{equation}
\rho_\text{vac}(q_0)=
\epsilon(q_0) -q_0\;\epsilon^\prime(q_0)=0\,.
\vspace*{-0mm} 
\label{rhovacq0}
\end{equation}
\end{subequations}
Even though both terms in the middle expression of \eqref{rhovacq0}
can be of order $(E_{\rm Planck})^4$, they cancel exactly
in the equilibrium state corresponding to Minkowski spacetime.
As such, \mbox{$q$--theory} provides a
possible solution~\cite{KlinkhamerVolovik2010}
of the \emph{main} cosmological constant
problem~\cite{Weinberg1988}, namely, why the energy scale
of the observed cosmological constant
is essentially zero compared to the known energy scales of
elementary particle physics (e.g., $E_{\rm QCD}\sim 10^2\;\text{MeV}$
and $E_{\rm ew} \sim 1\;\text{TeV}$).

\vspace*{-2mm} 
\subsection{Dynamics}
\label{subsec:Resume-Dynamics}

For a particular realization of the $q$ variable
and a nontrivial $q$--dependence of the gravitational coupling parameter,
it has been shown~\cite{KlinkhamerVolovik2008b}
that $q$ becomes spacetime-dependent
and so does the vacuum energy density
$\rho_\text{vac}(q)=\textstyle{\frac{1}{2}}\,(q-q_0)^2
+ \text{O}\big( (q-q_0)^3 \big)$.

In a spatially flat, isotropic, and homogeneous Robertson--Walker
universe, the field equations then give a rapidly oscillating
vacuum energy density $\rho_\text{vac}(t)$, which drops from a
Planck-scale value to zero as $t \to \infty$. Specifically,
the following behavior has been found~\cite{KlinkhamerVolovik2008b}:
\vspace*{-2mm} 
\begin{equation}
q(\tau)/q_0 -1 \sim \tau^{-1}\; \sin \tau \,,\quad
r_\text{vac}(\tau) \sim \tau^{-2}\; \sin^2 \tau\,,
\vspace*{-2mm} 
\label{rhoV-decay}
\end{equation}
where the dimensionless cosmic time $\tau$ and
the dimensionless vacuum energy density $r_\text{vac}$
have been obtained by appropriate scalings with microscopic
(Planckian) quantities such as $q_0$.

\section{Neutrino mass and effective momentum cutoff}
\label{sec:Neutrino-mass-cutoff}

In this appendix, a heuristic argument is given for a low effective
momentum cutoff of the mass effects from neutrino fields
in an interacting system
(in fact, the system relevant to the virtual particles
populating the quantum vacuum up to Planck-scale energies).

Start from the see-saw mechanism  for a single neutrino flavor
(see, e.g., Ref.~\cite{King2003} and references therein),
which gives a light neutrino mass, $M_\nu(0) = (M_{D})^2/M_{R}$,
for a right-handed Majorana scale $M_{R}$
and a Dirac scale $M_{D} \sim E_\text{ew}$ with hierarchy
$M_{R} \gg M_{D}$.

Adding linear momentum contributions to the standard see-saw mass matrix
gives the following effective mass matrix:
\vspace*{-2mm} 
\begin{equation}\label{eq:neutrino-matrix}
\mathcal{M}_\nu^\text{\,(lin)} (p^2)=
\left(
  \begin{array}{cc}
    c_{11}\,p & M_{D} + c_{12}\,p \\
    M_{D} + c_{12}\,p & M_{R} + c_{22}\,p\\
  \end{array}
\right)\,,
\vspace*{-2mm} 
\end{equation}
with $p\equiv \sqrt{p^2} \geq 0$ for a Euclidean metric
[see below \eqref{Dirac-propagator}]
and nonnegative numerical coefficients $c_{ij}$ of order $1$.
The $c_{11}$ term in \eqref{eq:neutrino-matrix}
corresponds to an effective Higgs-triplet contribution
from the interactions, independent from mass terms.
One eigenvalue of \eqref{eq:neutrino-matrix} is of order $M_{R}$.
The absolute value of the other eigenvalue is
\vspace*{-2mm} 
\begin{equation}\label{eq:Mnup}
M_\nu(p^2) = (M_{D})^2/M_{R} - c_{11}\,p + \text{O}(p^2)\,,
\vspace*{-2mm} 
\end{equation}
for $p \ll M_{D} \ll M_{R}$.
Observe that the two terms on the right-hand side of
\eqref{eq:Mnup} have a relative minus sign, which is
not the case for the individual entries of the
matrix \eqref{eq:neutrino-matrix}.

Equation \eqref{eq:Mnup} for $c_{11}\sim 1$ shows an effective
momentum cutoff for mass effects of the order of the neutrino
mass itself, as discussed in the fourth paragraph of
Sec.~\ref{sec:Fermion propagator}.
For the simple example considered in this appendix
and small enough $p^2$, the \textit{Ansatz} for the mass function
would be nonanalytic in $p^2$, namely,
$M_\nu(p^2) = \alpha/(\beta+p)$ with $\alpha \sim \beta^2$.


\end{document}